\documentclass{JHEP}
\usepackage{graphicx}
\usepackage{amsmath,amstext,amssymb}
\title{Weakly Warped Extra Dimensions and SN1987A}
\author{Patrick J. Fox\\ Department of Physics, Box 351560, University of Washington,
\\ Seattle, WA 98195-1560, USA \\ \email{pjfox@phys.washington.edu}}
\abstract{The neutrino pulse from SN1987A provides one of the most rigourous constraints
on models of extra dimensions.  Previously, calculations have been done to bound
the size of these extra dimensions in the case when the metric was
factorizable.  Here we consider the case of 2 `weakly warped' extra dimensions.
We find that even though weak warping seems only to affect the zero mode this
can have a measurable effect on the supernovae bounds.  In any braneworld model
such warping is necessarily present and as such should be taken into account in
supernovae bounds and in searches for corrections to Newtonian gravity.}
\preprint{UW-PT-00-19}

\begin{document}
\section{Introduction}

Recently there has been an explosion of interest in theories with extra 
spacetime dimensions as possible solutions to the gauge hierarchy problem 
and the cosmological constant problem, see for example
\cite{Arkani-Hamed:1998rs}, \cite{Antoniadis:1998ig}, \cite{Randall:1999ee}, \cite{Randall:1999vf}.  
These models can be tested in various ways depending on the size and warping of
the extra dimensions.  Such tests include, observing corrections to a $1/r$
gravitational potential at short distances, observing missing $p_{T}$ or new
spin 2 resonances in high energy colliders or measuring the shape of the
neutrino pulse from supernovae.

For the case of 2 extra dimensions, perhaps the most rigourous constraint comes
from the neutrino pulse observed  from SN1987A.  Various analyses have been
carried out for the case of toroidally 
compactified extra dimensions \cite{Hanhart:2000er}, \cite{Cullen:1999hc},
\cite{Barger:1999jf}.  However, if the Standard Model fields are constrained 
to lie on a 4 dimensional brane in the higher dimensional spacetime then the
energy density situated on the brane will cause a warping in the higher
dimensions.  The  compactification will no longer be toroidal but instead will be some
curved, compact manifold.  Therefore, we consider a weakly warped
 compactification to investigate how the warping affects the bounds coming from the 
supernovae data.

In order to do this we consider a model similar to that of Chen, Luty and 
Ponton \cite{Chen:2000at}. This model has two extra dimensions, which is the
most interesting phenomenologically.  We are interested in the effects of the 
light KK modes of the graviton on the neutrino pulse of SN1987A and we ignore the
radion since it has no KK modes \cite{Charmousis:2000rg}.  The supernova bounds
are sensitive to all the low lying KK modes (those whose energy is approximately
less than the temperature in the core of the star).  As such the effect of any one
particular mode coupling with gravitational strength is negligible. 

The strength of a KK mode's coupling to matter is determined by the ratio of
their wavefunction on the 3-brane to that of the zero mode.  All the KK modes
whose energy is greater than the warping scale will be unaffected by the
warping.  Thus, one might expect very little effect, for weak warping, since very
few modes are directly affected.  However, introducing a warping introduces a
potential for the zero mode graviton \cite{Randall:1999vf}, giving its
wavefunction a profile in the extra dimensions.  This results in the
normalization for the zero mode changing.  

The normalization of the zero mode sets the scale of gravitational interactions,
$G_{N}=N_{0}^{2}$,  which in turn affects the strength of the KK couplings.  A
KK mode's coupling to matter on our brane is determined by the mode's value at
our brane.  The zero mode's coupling to our brane determines the strength of
gravity.  So the ratio of a KK mode's value at the origin to that of the zero
mode determines the mode's coupling to matter relative to the strength of
gravity.  Thus, if the KK modes are unaffected by warping and $N_{0}$ becomes
smaller then the KK modes actually now couple with greater than gravitational
strength.  The effect of weak warping can be modeled by a tower of KK
modes whose masses don't change from the unwarped case but whose couplings are all shifted.

We find that, for the model considered here, the zero mode's coupling is
decreased by warping and so the effect of the KK modes increases\footnote{It is also
conceivable to have models where the warping increases the zero mode coupling,
for instance in models where we live at a maximum of the warp factor rather than
a minimum.}.  This results in a change in the supernova bounds between the
(unnatural) unwarped case and the warped case.  Notice that the effect of the
warping is to change the coupling strength of the KK modes from
that expected for the unwarped case.  In particular this means that in searches
for short distance corrections to gravity one should not expect the coupling of
the lightest KK mode, $\alpha$, to be simply determined by the number or
geometry of the extra dimensions, \cite{Kehagias:2000my}, \cite{Hoyle:2000cv}.

This paper is organized as follows.  In section 2 we describe the model and
solve the equations of motion for the KK modes of the graviton.  In section 3
we describe how the KK modes affect the neutrino pulse from SN1987A.  In section
4 we discuss the case of stronger warping and we conclude in Section 5.

\section{The Model}

We consider a variation of the (6 dimensional) `spaceneedle metric' of Chacko 
and Nelson \cite{Chacko:2000eb}.  The inner 4-brane on which the standard
model fields live is moved into the origin, and becomes a 3-brane, and we impose
a $Z_{2}$ symmetry around the outer 4-brane so that the radial extra dimension
ends again on an identical copy of the original 3-brane.  Such a scenario has
 also been considered by Chen, Luty and Ponton \cite{Chen:2000at}.
Explicitly we consider a metric of the form
\begin{equation}
ds^{2}=f(r)\eta_{\mu\nu}dx^{\mu}dx^{\nu}+s(r)d\theta^{2}+dr^{2},
\end{equation}
where $\mu, \nu$ run over our 4 spacetime dimensions (we take a `mostly plus
metric')  and $\theta, r$ span the 2 extra dimensions.

The stress energy on the 3-brane is given by\footnote{Following notation of Chen, 
Luty and Ponton} $T_{ab}=diag(T_{4},T_{4},T_{4},T_{4},0,0)$ and on the 4-brane by
 $T_{ab}=diag(T_{5},T_{5},T_{5},T_{5},T_{5,\theta},0)$.  This model has a 
global deficit angle that is proportional to the 3-brane tension and there is
one overall fine tuning that is necessary to make the 3-brane flat.  

Solving Einstein's equations for this metric and matching across the 
boundaries at the branes produces a solution of the form, 
\begin{equation}
f=\cosh^{4/5}(\frac{5}{2}\sqrt{k}r),\quad s=\frac{{f'}^{2}}{f},
\end{equation}
where $k=-\Lambda_{6}/10M_{6}^{4}$ with $\Lambda_{6}$ being the bulk 
cosmological constant which we will take to be negative and $M_{6}$ 
the fundamental Planck scale.

We are interested in the properties for the graviton KK modes and as such will
need to solve the equations of motion for a perturbation about this solution.
Consider a perturbation of the form,
\begin{equation}
ds^{2}=(f(r)\eta_{\mu\nu}+h_{\mu\nu}(r,x^{\alpha}))dx^{\mu}dx^{\nu}+s(r)d\theta^{2}+dr^{2}.
\end{equation}
We expand $h_{\mu\nu}(r,x^{\alpha})$ in plane waves,
\begin{equation}
h_{\mu\nu}(r,x^{\alpha})=\sum_{p}e^{ip.x}h_{p}(r),
\end{equation}
and using Einstein's equations we find that $h$ (where we henceforth drop the subscript
$p$ for convenience) obeys the equation of motion,
\begin{equation}\label{eq:gravitoneom}
-h''-\frac{1}{2}\frac{s'}{s}h'-\left(\frac{f'}{f}\right)^{2}h+10kh=\frac{m^{2}}{f}h,
\end{equation}
where $p^{2}=\eta_{\alpha\beta}p^{\alpha}p^{\beta}=-m^{2}$.   We are ignoring
modes that have  $\theta$ dependence since they 
couple to momentum in the $\theta$ direction and thus they do not couple to particles living on 
the 3-brane.  Such modes can be pair produced by SM 
particles on the brane but these processes will be suppressed.
The boundary conditions imposed on $h$ by the presence of the branes at $r=0$
and $r=L$ respectively are,
\begin{align}
h'(0)=0 & \text{ and } \Delta \frac{h'}{h}=-\frac{T_{5,\theta}}{2}.
\end{align} 
As it stands (\ref{eq:gravitoneom}) is difficult to solve  so we make the
simplifying assumption that the warping is weak, i.e. $\sqrt{k}L<1$.  We will
only keep the dominant terms in an expansion in $\sqrt{k}L$.  Upon solving this
simplified equation of motion we will see that the solution is entirely
consistent with the assumptions we have made.  For weak warping the
(\ref{eq:gravitoneom}) becomes,
\begin{equation}
-h''-\frac{1}{r}h'+10kh=m^{2}h.
\end{equation}
The solution, after imposing the boundary condition at the 3-brane, is
\begin{equation}
h=N_{m}J_{0}\left(\sqrt{m^2-10k}r\right).
\end{equation}
Where the allowed values of $m^2$ will be specified by applying the boundary
conditions at the 4-brane.  For weak warping this is,
\begin{equation}
\left.\frac{dJ_{0}/dr}{J_{0}}\right|_{r=L}=5kL-\frac{125}{12}k^{2}L^{3}+O(k^{3}L^{5}).
\end{equation}
This results in the masses being given by,
\begin{equation}
m_{n}^2=\left(\frac{\nu_{1n}}{L}\right)^{2}+\frac{125}{6}k^{2}L^{2}+O(k^{3}L^{4}).
\end{equation}
Where $\nu_{1n}$ is the nth zero of $J_{1}$; for large masses the spacing of
these zeros becomes $\pi$.  

In order to find N we must normalize the wave functions for the KK modes i.e.,
\begin{align}
N_{n}^{-2}=& \int_{0}^{\theta_{max}}d\theta\int_{0}^{L}rdr
\left|J_{0}(\nu+\delta)(\frac{r}{L}))\right|^{2}\\
=& \theta_{max}L^{2}\int_{0}^{1}dz z \left|J_{0}((\nu+\delta)z)\right|^{2}\\
=& \frac{\theta_{max}L^{2}}{(1+\delta ')^{2}}\int_{0}^{1+\delta '}dw w
\left|J_{0}(\nu w)\right|^2 \text{ where } \delta '=\delta/\nu,
\end{align}
where $\delta=-5kL^{2}/\nu  + O(k^{2}L^{4})$.
Now, $\int_{0}^{1+\delta '}dw w J_{0}^{2}=F(\delta ')$ where we can then expand
$F(\delta ')$ in a  Taylor series about $\delta '=0$.  We find that,
\begin{equation}
N_{n}^{-2}=L^{2}\theta_{max}J_{0}^{2}(\nu_{1n})\left\{\frac{1}{2}+2{\delta
'}^{2}+O({\delta '}^{3})\right\}.
\end{equation}
We can carry out a similar calculation for the zero mode and we find that the
normalization $N_{0}$ is given by,
\begin{equation}
N_{0}^{2}=\frac{2}{\theta_{max}L^{2}}\left[1-\frac{5}{2}L^{2}k\right].
\end{equation}

From the above we see that, to leading order in $kL^{2}$, the only effect of the
warping is on the normalization of the zero mode, the spacing and
normalization of the higher modes are unaffected.  Thus, as mentioned above,
this now means that the KK modes couple with greater than gravitational strength
since it is the KK mode to zero mode ratio that determines the strength of
coupling.  The density of states of the KK modes is unchanged.

Intuitively, when we introduce warping the quantum mechanics problem that
corresponds to the equation of motion of the graviton acquires a potential that
is no longer flat but has a `volcano potential' as in the RS models.  Now since
in our model the graviton zero mode is peaked around the other brane (an effect
of the volcano potential) and the KK modes are orthogonal to the zero mode the
KK modes will be stronger at our brane than in the case with no warping.  Thus
they interact more strongly.  Conversely, if we had been living at a maximum of
the warp factor the coupling of the KK modes would be decreased by warping.  The
effect of light warping can be modeled simply by altering the coupling strength
of the tower of KK modes of the graviton, keeping their spacing the same as the
unwarped case.  This allows us to easily adapt results derived for flat
compactifications.  Let us calculate the effect on supernova bounds.

\section{Bounds}

A theory with compactified extra dimensions can be thought of in 2 different ways
depending on the energy regime we are interested in.  If only the first few KK
modes can be excited then the theory can be thought of from a 4D point of view.
We have the massless graviton and a tower of KK modes.  Varying the warping and
brane spacing shifts the spacing and coupling of the KK modes.  If we are at a
sufficiently high energy such that many of the KK modes can be excited it makes
sense to think of the theory in its fundamental 6D description.

From the above we saw the effect of weak warping was to raise the coupling of
the KK modes without shifting their masses.  This means that in short distance
gravity experiments where only the first few KK modes are accessible the strength
of these modes are not simply related to the geometry of the compactification
\cite{Kehagias:2000my}, \cite{Hoyle:2000cv}.  Instead the warping that is
necessarily present in braneworld models changes the coupling of these modes
from that expected due to geometry alone.

However, in a supernova the temperature is on the order of 30 MeV.  For weak
warping many KK modes are excited, so the theory is best viewed from a 6D point
of view.  The effects of the KK tower have to be summed up and result in a
correction to Newtonian gravity,
\begin{equation}\label{eq:newtoniancorrections}
F_{Newton}\rightarrow \frac{m_{1}m_{2}}{M_{pl}^{2}r^{2}}\left(
  1+\frac{M_{pl}^{2}}{M_{*}^{4}}\frac{1}{r^{2}}\right).
\end{equation}

The basic principle used to place supernova bounds on the sizes and warpings of the extra
dimensions is the idea that having extra light modes to excite in supernovae
explosions could change the shape of the neutrino pulse associated with these explosions.
Since we know the shape of the neutrino pulse from SN1987A reasonably well any
new physics that would significantly alter the pulse from that observed can be
ruled out.  The extra light modes present (with masses up to the average energy in
the proto-neutron star) could be excited and thus would `steal' energy from the
neutrinos.  A rule of thumb is that if the emissivity of this new energy loss
processes is higher than $10^{19} ergs/g/s$ then the neutrino signal would be
sufficiently different from that observed that we would be able to rule the
processes out \cite{Raffelt:1996wa}.

The dominant processes, contributing to this energy loss and involving the graviton,
is bremsstrahlung of KK gravitons from nucleons, $NN\rightarrow NNh$.  Hanhart
\emph{et al.} \cite{Hanhart:2000er} computed these processes in a model
independent way, relating their emissivities to measured nucleon-nucleon cross
sections.  However, in their calculation the extra dimensions were compactified
on a flat torus.   We can easily adapt their calculation to the case of weak
warping.  There are two changes that need to be made to their calculation, 
\begin{enumerate} 
\item We have warping in our extra dimensions \cite{Hanhart:2000er} does not.
\item  Hanhart \emph{et al.} compactified the extra dimensions on a torus,
  whereas we have compactified the extra dimensions on a cone\footnote{The extra
    dimensions are topologically equivalent to a sphere, they have a deficit
    angle singularity at the origin due to the 3 brane and a curvature singularity at the
    orbifold point due to the 4 brane.  In the limit of zero warping the extra
    dimensions are a `wedge' cut from a sphere.}.
\end{enumerate}

First the effect of warping, this changes the coupling of the KK modes.  In
\cite{Hanhart:2000er} the coupling is defined to be $\kappa_{h}=\sqrt{32\pi
  G_{N}}$ but for us it is $\kappa^{2}=\frac{32\pi
  G_{N}}{J_{0}^{2}(\nu_{1n})}(1+5/2kL^{2})\equiv \frac{32\pi G_{N}}{J_{0}^{2}(\nu_{1n})}g_{KK}^{2}$.  
Secondly, the change in geometry results in a change in the density of states of
the KK modes,
\begin{align}
L^{2}\Omega_{1}\int dm m &\rightarrow \frac{L}{\pi}g_{KK}^{2}\int
\frac{dm}{J_{0}^{2}(mL)}\\
& =\frac{1}{2}L^{2}g_{KK}^{2}\int m dm.
\end{align}
Where in the last step we have used the asymptotic form of the Bessel function,
$J_{0}^{2}(x)\sim 2/\pi x \cos^{2}(x-\pi/4)$ so $J_{0}^{2}(\nu_{1n})\sim 2/\pi mL$. 
$g_{KK}$, the KK mode coupling, can be varied independently of the brane spacing
and in the limit of zero warping it is 1.

Thus we can relate the flat results, $F(R)$, with a torus of radius $R$ to the
warped results, $W(k,L)$ with warping scale $k$ and radius $L$:
\begin{equation}\label{eq:emmissivity}
W(k,L)=F(R)*(R^{2}\Omega_{1})^{-1}g_{KK}^{2}\frac{L^{2}}{2},
\end{equation}
where $\Omega_{1}$ is the surface area of a unit 1-sphere.

In order to see the effects of the warping and the different geometry on the
results of Ref. \cite{Hanhart:2000er} we plot emissivity against temperature for
various models.  As in \cite{Hanhart:2000er} we take the density of nucleons in the
neutron star to be that of nuclear matter, $0.16 fm^{-3}$.  Fig
\ref{fig:simplechange} shows emissivity for the various models.  The dashed line
shows the emissivity for the model of \cite{Hanhart:2000er} which has flat
toroidal extra dimensions, the dot-dashed line is for the conical geometry with
no warping and the solid line is for the conical geometry with a warping of
$k=0.5mm^{-2}$, $R=L=1mm$ in all cases.   

\FIGURE{\includegraphics[scale=0.5,angle=270]{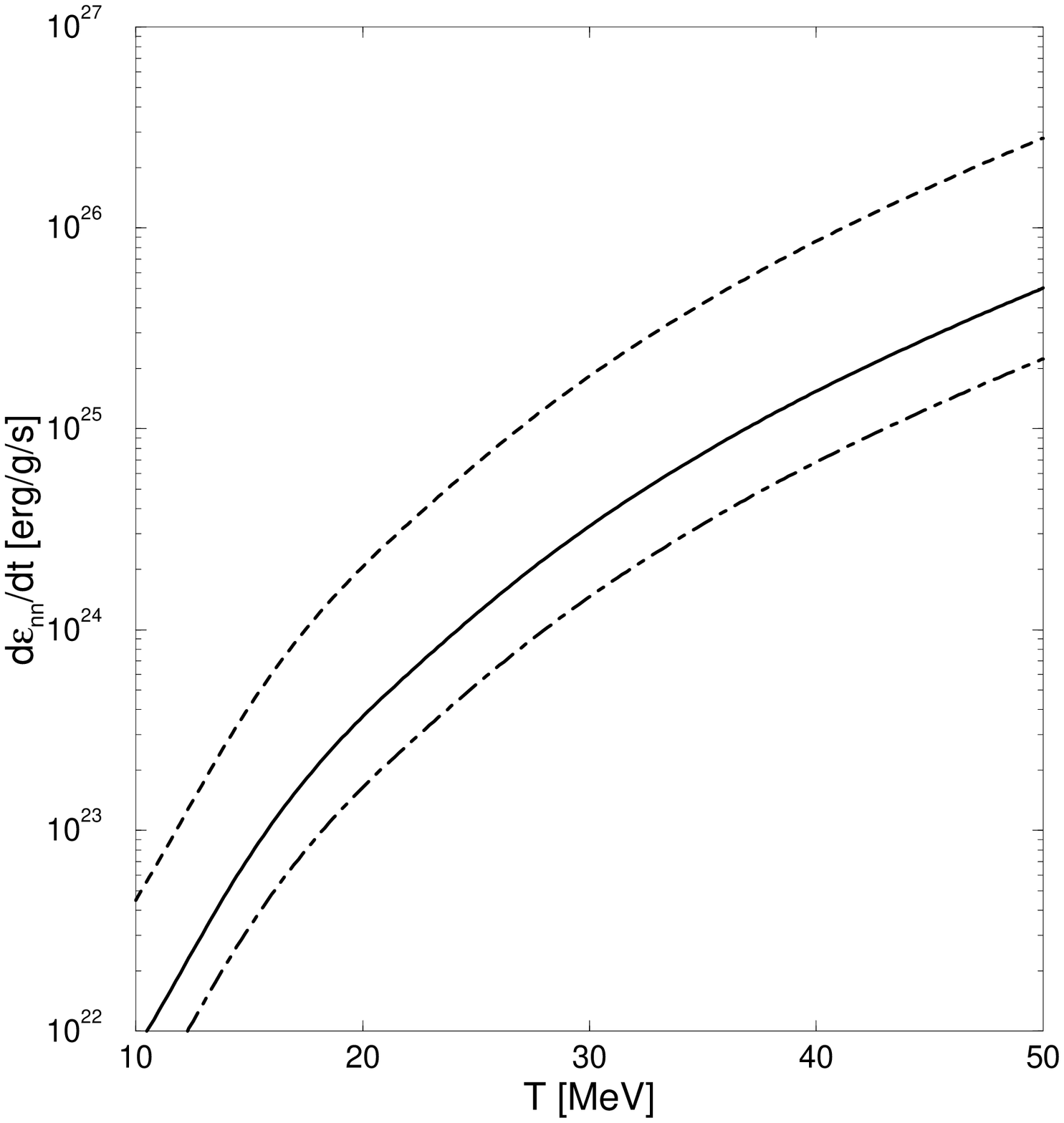}\caption{Emissivity
    due to graviton emission in the cases of flat toroidal compactification
    (dashed line), flat conical compactification (dot-dashed line) and warped
    conical compactification (solid line).\label{fig:simplechange}}}

Comparing the two similar geometries represented by the dot-dashed and the
solid lines we can see the effect of warping is to increase the energy
emission through graviton KK emission, as described earlier.  In going from the flat toroidal
compactification to the warped conical compactification there is also a change
in geometry.  This change in geometry overrides the effect of the warping and
results in the energy loss being greater in the flat case.  However, if two similar
geometries are considered the effect is determined by warping alone and
the energy loss increases.

(\ref{eq:newtoniancorrections}) can be derived by summing up the effect of all KK
mode exchange, so $M_{*}^{-4}\sim g_{KK}^{2}$.  Together with
(\ref{eq:emmissivity}) this tells us that the emissivity bound on extra
dimensions is an immediate bound on $M_{*}$.  

In a supernova the available energies are sufficiently high that it effectively
`sees' all of the KK modes and so experiences the true 6D theory.  Thus, for
weak warping, the supernova can place bounds on $M_{*}$ beyond those
already known from short distance gravity experiments.  In a warped model both the coupling and
the spacing are parameters, there are many combinations that result in the same
$M_{*}$.  For weak warping, with $k\approx 1/L^{2}$, the bounds on the brane
spacing for various supernova temperatures are:
\begin{align}
L\lesssim & 10^{-2} mm &\quad & T=20MeV \\
L\lesssim & 2 \times 10^{-3} mm &\quad &  T=30MeV\\
L\lesssim & 5 \times 10^{-4} mm &\quad & T=50MeV
\end{align}

\section{Stronger Warping}

So far we have been interested in the case $\sqrt{k}L<1$ but there are other regimes
with stronger warping that can also be considered.  The assumption $k\gg
1/L^{2}$ makes (\ref{eq:gravitoneom}) more tractable.  In this limit
(\ref{eq:gravitoneom}) becomes,
\begin{equation}
-h''-\sqrt{k}h'+6kh=2^{4/5}m^{2}e^{-\sqrt{k}r}h.
\end{equation}
This has a solution,
\begin{equation}
h=Ae^{-1/2\sqrt{k}r}\left(BJ_{5/2}\left(\frac{2^{2/5}me^{-\sqrt{k}r}}{\sqrt{k}}\right)+Y_{5/2}\left(\frac{2^{2/5}me^{-\sqrt{k}r}}{\sqrt{k}}\right)\right),
\end{equation}
A and B are constants to be determined by boundary conditions.  

Calculating the mass and couplings of the KK modes for this solution becomes a
difficult problem.  It is necessary to match at the boundaries and to normalize
the wavefunction to find A, B and the allowed masses.  We have not been able to
find an analytic closed form for the emissivity in this regime.  To do so would
require a numerical investigation and as such is beyond the scope of this
paper.  For the low lying KK modes whose masses are small enough that the
small argument limit of the Bessel functions can be used we find that the
boundary condition at $r=L$ implies $B\approx
3\left(\frac{2^{2/5}me^{-\sqrt{k}L}}{\sqrt{k}}\right)$.  Thus the coupling
strength of the KK modes becomes energy dependent.  The effects of warping are
no longer determined by the zero mode alone. 

If $k<MeV^{2}$ the majority of the KK modes excited in the supernova are unaffected by
the warping.  The emission of KK modes with $m\gg \sqrt{k}$ can be thought of as bremsstrahlung from
nucleons into a 6 dimensional spacetime.  In this fundamental 6D picture the
gravitational coupling strength of the KK modes is determined by $M_{*}$.  The
energy loss to KK emission is determined by $M_{*}$.  Although the supernova
data could not be used to bound $L$ or $k$ separately, it can be used to place
bounds on $M_{*}$.  The bound on $M_{*}$ is essentially unchanged from that of
the flat case.

If $k\gg MeV^{2}$ then all the KK modes that will be excited in the supernova
will feel the warping.  This scenario is more like that of
Ref. \cite{Randall:1999ee} and the supernova data is of little use in placing
bounds in such a situation. 

\section{Conclusions}

Warping is a necessary feature of braneworld scenarios and we have demonstrated
that even very weak warping can have a measurable affect on low energy physics.
For warping on the scale of the size of the extra dimensions, $k\sim 1/L^{2}$,
we find that the leading order effect is to change the coupling of the KK
modes.  The mode spacing is relatively unchanged since their masses are
$\gg L^{-1}$ so they are unaffected by the warping. 

This change in KK mode coupling affects bounds coming from SN1987A and short
distance gravity experiments.  For the particular model considered here we found
that this strengthens the bounds on $L$ from those of the unwarped case.  If we
were living at a local maximum of the warp factor we would expect the bounds on
$L$ to be weakened.  The latter case might be testable in short distance gravity
experiments.  In either case we expect the fundamental scale to be too high to
be probed by accelerators.

\acknowledgments

I would like to thank Ann E. Nelson for many very helpful discussions.  I also
thank Zacharia Chacko and Christoph Hanhart for useful conversations.

\end{document}